\documentclass[conference]{IEEEtran}
\IEEEoverridecommandlockouts
\usepackage{cite}
\usepackage{amsmath,amssymb,amsfonts}
\usepackage{algorithmic}
\usepackage{graphicx}
\usepackage{textcomp}
\usepackage{xcolor}
\def\BibTeX{{\rm B\kern-.05em{\sc i\kern-.025em b}\kern-.08em
    T\kern-.1667em\lower.7ex\hbox{E}\kern-.125emX}}
\begin{document}
\vspace{-16mm}
\title{Automatic Displacement and Vibration Measurement in Laboratory Experiments with A Deep Learning Method\\
% {\footnotesize \textsuperscript{*}Note: Sub-titles are not captured in Xplore and
% should not be used}
% \thanks{the U.S. National Science Foundation}
}
\vspace{-10mm}
\author{\IEEEauthorblockN{Yongsheng Bai, Ramzi M. Abduallah, Halil Sezen, Alper Yilmaz}
\IEEEauthorblockA{\textit{Department of Civil, Environmental and Geodetic Engineering} \\
\textit{The Ohio State University}\\
Columbus, OH, USA \\
bai.426@osu.edu, abduallah.2@osu.edu, sezen.1@osu.edu, yilmaz.15@osu.edu}
\vspace{-10mm}
% \and
% \IEEEauthorblockN{2\textsuperscript{nd} }
% \IEEEauthorblockA{\textit{Dep. Civil, Environmental and Geodetic Engineering} \\
% \textit{The Ohio State University}\\
% Columbus, OH, USA \\
% abduallah.2@osu.edu}
% \and
% \IEEEauthorblockN{3\textsuperscript{rd} Halil Sezen}
% \IEEEauthorblockA{\textit{Dep. Civil, Environmental and Geodetic Engineering} \\
% \textit{The Ohio State University}\\
% Columbus, OH, USA \\
% sezen.1@osu.edu}
% \and
% \IEEEauthorblockN{4\textsuperscript{th} Alper Yilmaz}
% \IEEEauthorblockA{\textit{Dep. Civil, Environmental and Geodetic Engineering} \\
% \textit{The Ohio State University}\\
% Columbus, OH, USA \\
% yilmaz.15@osu.edu}
}

\maketitle
\begin{abstract}
This paper proposes a pipeline to automatically track and measure displacement and vibration of structural specimens during laboratory experiments. The latest Mask Regional Convolutional Neural Network (Mask R-CNN) can locate the targets and monitor their movement from videos recorded by a stationary camera. To improve precision and remove the noise, techniques such as Scale-invariant Feature Transform (SIFT) and various filters for signal processing are included. Experiments on three small-scale reinforced concrete beams and a shaking table test are utilized to verify the proposed method. Results show that the proposed deep learning method can achieve the goal to automatically and precisely measure the motion of tested structural members during laboratory experiments.      
\end{abstract}

\vspace{-2mm}
\section{Introduction}
\vspace{-1mm}
For safety assessment and structural health monitoring of infrastructures such as buildings and bridges and their components, their vibrations and deformations need to be recorded and evaluated. Traditional structural sensors such as Linear Variable Displacement Transducers (LVDTs), dial gauges, accelerometers and other advanced sensors are used to measure deformations and vibrations. However, in some cases, conventional sensors may not be a good option to access the desirable instrumentation locations and work in a timely and cost-efficient way. Most importantly, traditional sensors measure the displacement or vibration at a discrete location, i.e., one sensor is used to measure one quantity at a single point. On the other hand, a high definition video camera can record the movement of a component or an entire structure rather than a single point.  

Recent developments on vision- and vibration-based technologies led to the measurement applications of low-cost and non-contact sensors on deformation and vibration monitoring of the infrastructures \cite{dong2020review}, especially because unusual and extreme deformations or vibrations in aging bridges and buildings may be an indication of significant serviceability or safety issues. Cameras can collect high-quality images or high-speed videos in lab or field tests as non-contact and non-destructive sensors. With computer vision technologies and various deep learning techniques, the vision sensors can not only be used as eyes for Artificial Intelligence (AI) vehicles and machines, but also provide opportunities for scientific measurements to researchers and engineers. In a laboratory experiment, cameras can be fixed near the testing station and record live motion of the monitored structural members. They can be placed at a stationary location or some distance away from an in-service bridge or building to record its structural movements remotely in the field. The collected visual data can be processed further to identify potential damage and assess the motion of the observed structures precisely.

To better understand and efficiently use displacement and vibration measurement data from cameras, we used Mask R-CNN \cite{he2017mask} with High-resolution network (HRNet) \cite{sun2019high} to track a target attached on beam specimens in the laboratory and to measure their deflections in the first study. Then the Mask R-CNN was applied on a shaking table test to track the dynamic motion of four targets simultaneously in the second study.
\vspace{-1mm}
\section{Literature Review}
Computer Vision (CV) and deep learning techniques are very useful to gain high-level understanding and extraction of desired information and precise motion measurements from images and videos. 

Traditional CV techniques have been widely used by researchers for displacement or vibration measurements by cameras. These techniques include image processing technique \cite{lee2006real}, up-sampled cross correlation \cite{feng2015vision}, adaptive Region of Interest algorithm  \cite{lee2017computer}, modified Taylor approximation \cite{liu2016vision}, and contour extraction with Speeded-Up Robust Features (SURF) \cite{yin2014concrete}. In addition, Lucas-Kanade template tracking algorithm \cite{guo2016dynamic, Dong2019mark-free} and Digital Image Correlation (DIC) \cite{chen2021homography} are employed to track displacement and vibration of structural members. Furthermore, Hu and Pai \cite{doi:10.1061/(ASCE)EM.1943-7889.0000374} utilized a camera-based 3D motion analysis system to measure the resonant vibration of steel cables. Chen et al. \cite{chen2017video} described an application with a video camera-based technique to test the vibration of an antenna tower on a tall building when a camera was placed 175 meters far from it. Hoskere et al. \cite{hoskere2019vision} used an Unmanned Aerial Vehicle (UAV) to measure the modal properties and dynamic response of a full-scale structure.

Deep learning is a relatively new research area for visual measurement applications. Dong et al. \cite{dong2020structural} implemented a full field optical flow algorithm named FlowNet2 to measure the displacement and vibration of structures. They also used the Spatio-Temporal Context Learning to track targets and utilized a Taylor approximation to gain subpixel level precision for displacement measurement \cite{dong2019robust}. Also, Dong et al. \cite{dong2019non} applied Visual Graph Visual Geometry Group (VGG) to extract features of the target for monitoring and measuring during the traffic time. These methods indicate how we can efficiently use cameras to monitor and measure the displacement and vibration of structural members or a structure in the laboratory or in the field.   
\vspace{-4mm}
\section{Methodology}
\begin{figure}[h!] % h:here, t:top
        \centering
        \vspace{-1mm}
        \includegraphics[width=\linewidth, height=2.4cm]{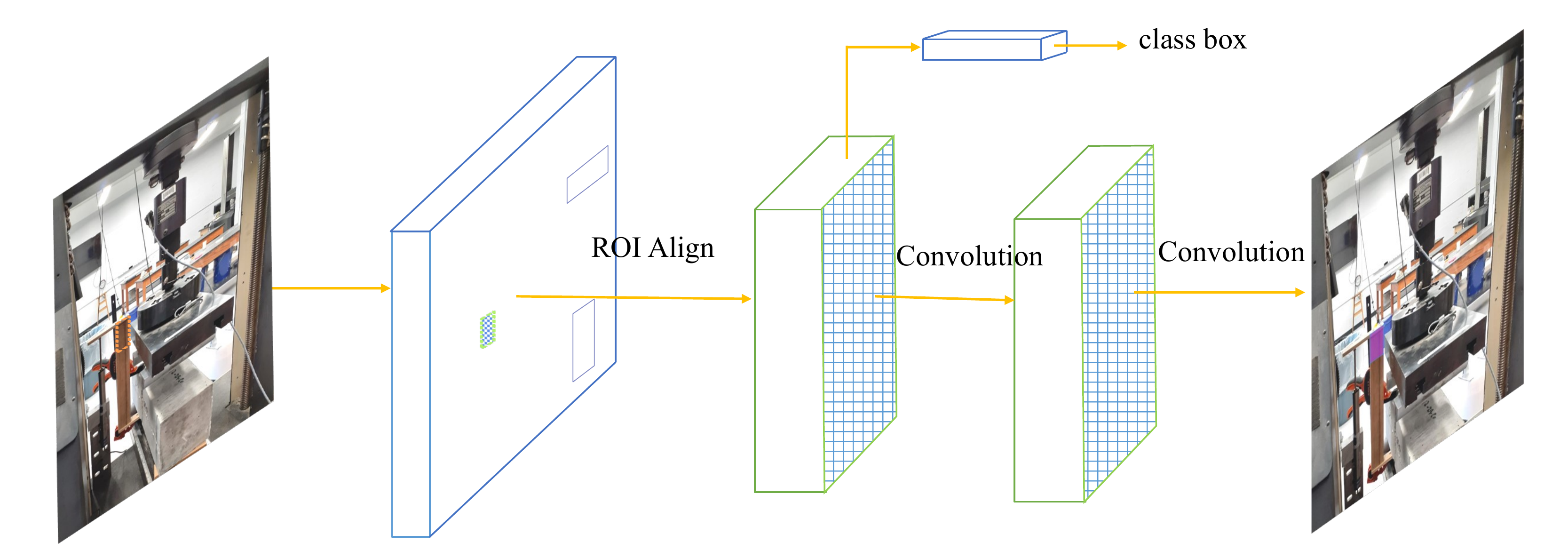}
        \vspace{-8mm}
        \caption{The framework of Mask R-CNN for tacking and measuring displacement of a reinforced concrete beam}
        \vspace{-8mm}
        \label{fig:maskrcnn}
\end{figure}

In our previous studies\cite{bai-2021-isprs, Bai1835end}, new variants of the latest Mask R-CNNs were successfully applied for structural damage detection with high accuracy. Therefore, we used one of the variants, Mask R-CNN with HRNet, to track and measure the displacement and vibration in this research. Fig. \ref{fig:maskrcnn} shows the framework of this Mask R-CNN.
\begin{figure}[h!] % h:here, t:top
        \centering
        \includegraphics[width=0.9\linewidth, height=3.2cm]{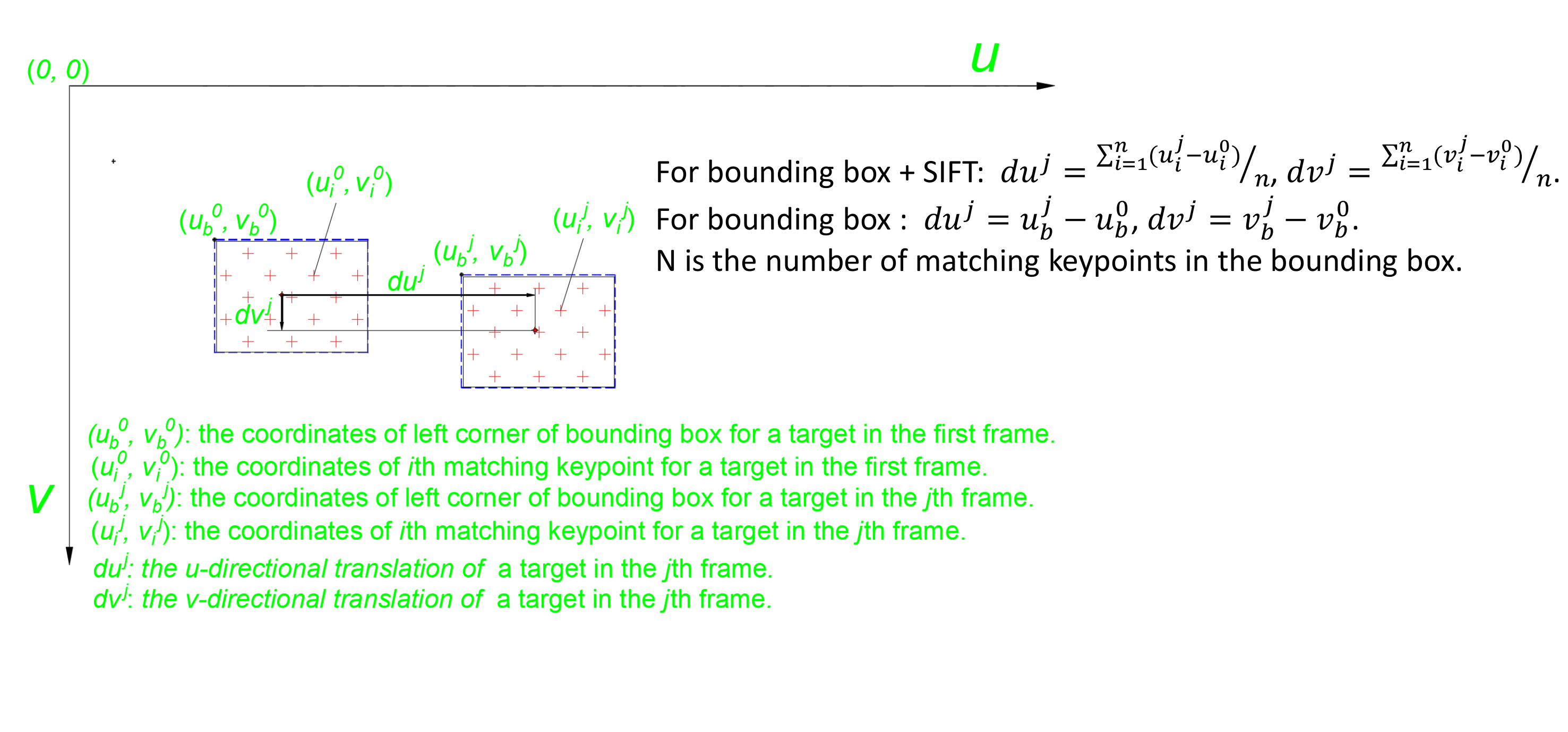}
        \vspace{-8mm}
        \caption{Translation of a target measured by a bounding box or by matching keypoints.}
        \vspace{-6mm}
        \label{fig:workflow}
\end{figure}
\begin{figure}[h!] % h:here, t:top
        \vspace{-2mm}
        \centering
        \includegraphics[width=0.9\linewidth, height=1.8cm]{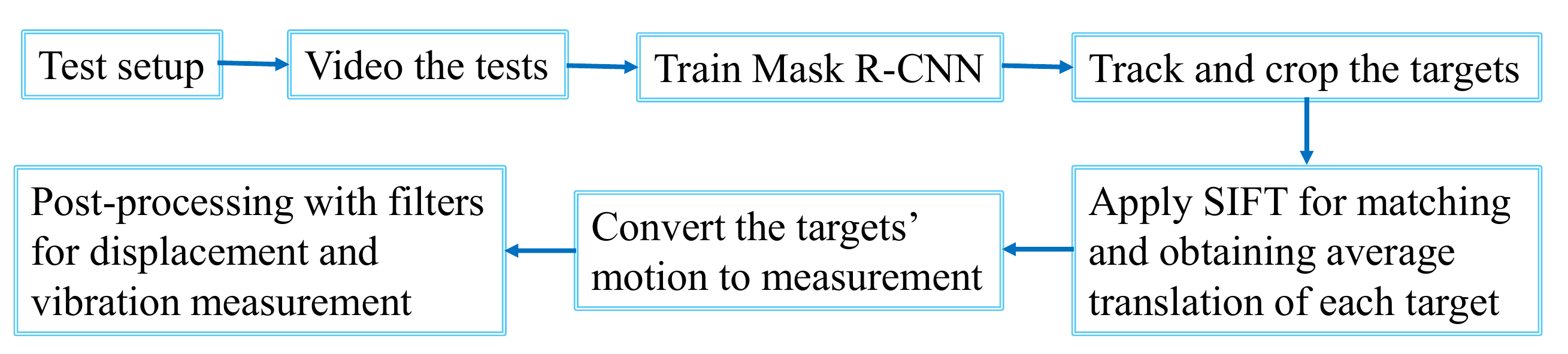}
        \vspace{-4mm}
        \caption{A flowchart of Mask R-CNN and SIFT for automated displacement and vibration measurement in a lab experiment with a stationary camera.} 
        \vspace{-4mm}
        \label{fig:flowchart}
\end{figure}

As shown in Fig. \ref{fig:maskrcnn} and \ref{fig:setup}, a wood frame is attached near the midspan of the tested beam so that it moves downward or upward when the beam is loaded or unloaded. The motion of the frame represents the deflection of the point where it is attached. Mask R-CNN is used to track the top of this wood frame (i.e., yellow dashed line on the left image in Fig. \ref{fig:maskrcnn}), which is marked by a bounding box and a mask in purple on the right image. Since the tracking target is a rigid body, its motion can be represented by any point on it or by the bounding box. On the other hand, as shown in Fig. \ref{fig:workflow} for the image plane of a stationary camera, translation of a target between the first frame and the $j$th frame, $du\textsuperscript{j}$ and $dv\textsuperscript{j}$, can be calculated as the position change of the bounding box or the average motion of the matching keypoints with SIFT. We observed that the mask for a target may be not exactly the same as its real shape in some cases. Therefore, SIFT is introduced into the pipeline to eliminate this inadvertent drawback when the bounding box is inaccurate to represent the target. Furthermore, subpixel precision can be achieved by matching and using the average motion of these keypoints on the target. In our pipeline (see Fig. \ref{fig:flowchart}), the ratio of good matching is restricted as SIFT commonly used in practice, but the range of coordinate change for each matching keypoint is also constrained such that the top good matching keypoints can be secured. Therefore, the mismatching is reduced dramatically and the accuracy of measurement is improved. Finally, the measurement is converted from pixel to length unit (inches or millimeters), which is also called a scalar, $s$. The horizontal and vertical displacement $dx\textsuperscript{j}$ and $dy\textsuperscript{j}$ of the target can be obtained as follows:
\vspace{-2mm}
\begin{equation}
{dx\textsuperscript{j}  = s \times du\textsuperscript{j} \label{eq1}}
\vspace{-2mm}
\end{equation}
\begin{equation}
{dy\textsuperscript{j}  = s \times dv\textsuperscript{j}\label{eq2}}
\vspace{-2mm}
\end{equation}  

As a comparison, an optical flow method called Lucas-Kanade (LK) tracker \cite{bouguet2001pyramidal} was used to track and measure the same targets in our experiments. For the LK tracker, the relative displacement is measured between two adjacent frames in a video. The final displacement is the sum-up of all the relative measurements. In addition, Savitzky-Golay filter \cite{bianchi2007electronic} and Butterworth filter \cite{press1990savitzky} were employed to handle the inconsistency and noise of the measurements. Fast Fourier Transform (FFT) \cite{cooley1965algorithm} was applied to extract the frequencies of the vibrating targets. 
\vspace{-2mm}
\section{Implementation}
\begin{figure}[h!] % h:here, t:top
        \centering
        \includegraphics[width=0.6\linewidth, height=2.4cm]{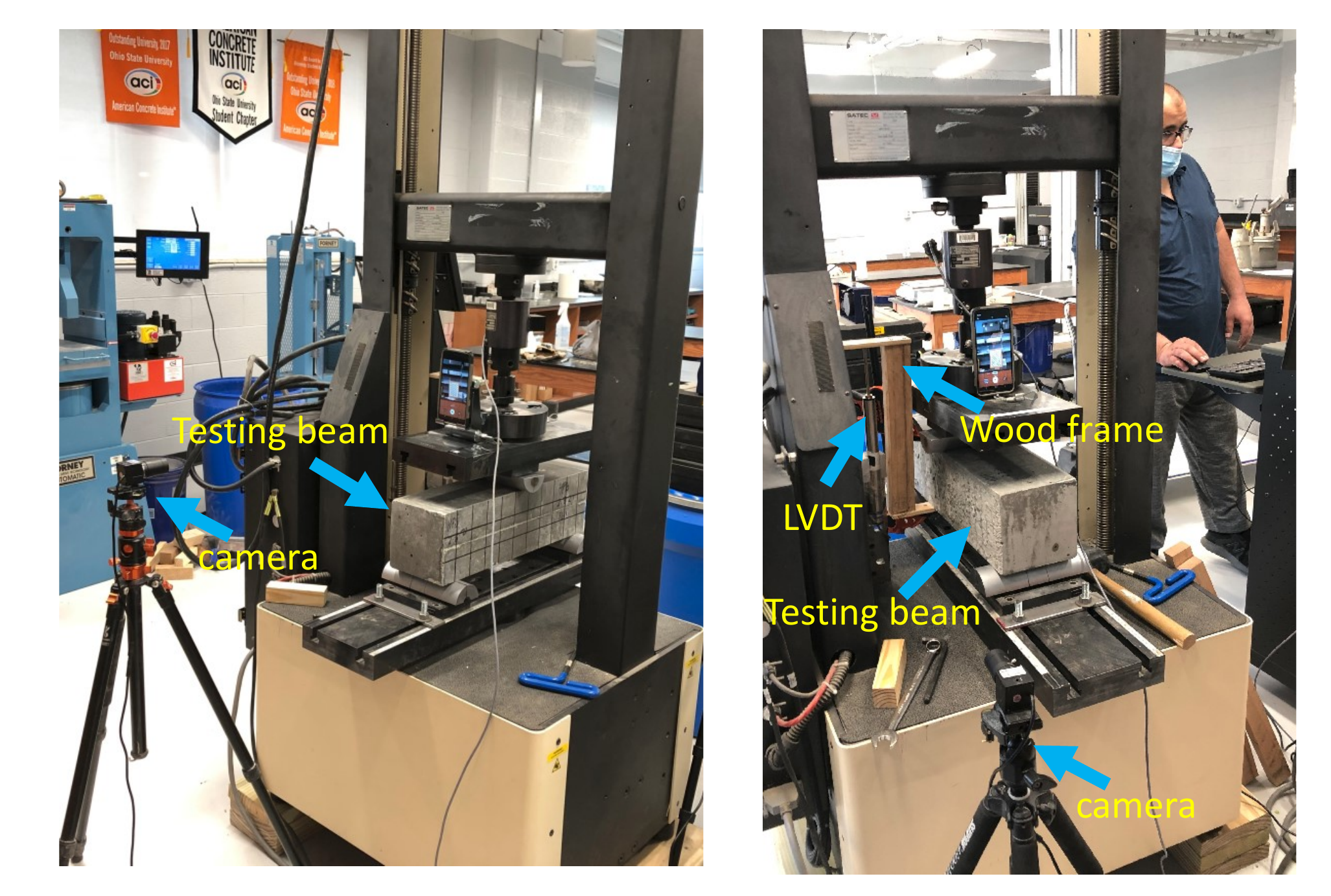}
        \vspace{-2mm}
        \caption{A flexural test of a RC beam with a LVDT and a camera.} 
        \label{fig:setup}
        \vspace{-4mm}
\end{figure}

\begin{figure}{} % h:here, t:top
        \centering
        \vspace{-4mm}
        \includegraphics[width=0.8\linewidth, height=2.4cm]{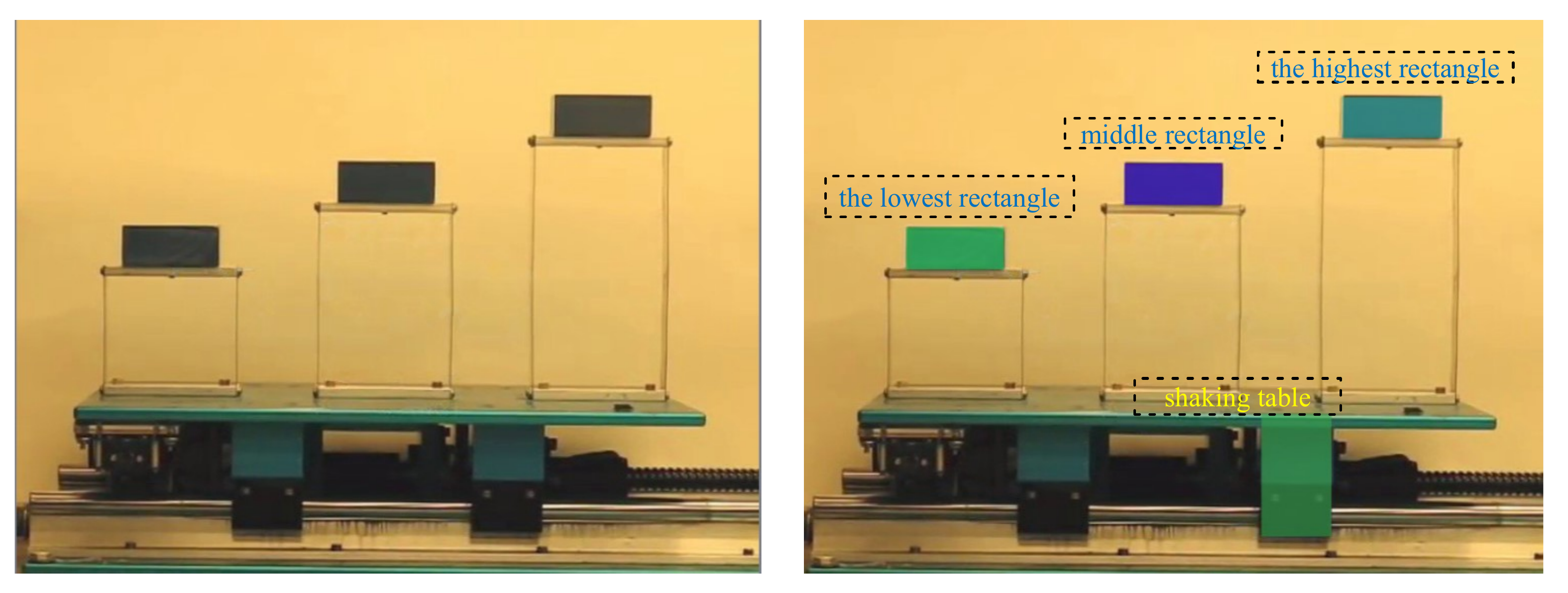}
        \vspace{-2mm}
        \caption{Original image (left) and label (right) in the shaking table video \cite{quanser_2017}.} 
        \label{fig:shaking}
\end{figure}

Two types of indoor experiments were utilized to verify our proposed methods. The first one is a small-scale experiment involving three reinforced concrete (RC) beams (see Fig. \ref{fig:setup}) conducted on the main campus of The Ohio State University in Columbus, Ohio. These beams are loaded and deflected until failure. Another test is an application on a video of a shaking table test (see Fig. \ref{fig:shaking}) \cite{quanser_2017}. 
\vspace{-4mm}
\vspace{-2mm}
\subsection{Deflection Measurement of RC Beams in Laboratory Tests}
In this experiment, three RC beams were subjected to a monotonically increasing point load at the midspan. A displacement sensor (a LVDT or dial gauge) and a wood frame were used to measure the deflection near midspan of the beams, while a camera was placed 3-feet away from the midspan (see Fig. \ref{fig:setup}). Its definition was $1600\times1200$ and the frame rate was 15 per second. The wood frame was clamped on the beam to represent the deflection of the targeted point on the tested beam.  

\begin{figure*}[h!] % h:here, t:top
        \centering
        \includegraphics[width=1\linewidth, height=3.cm]{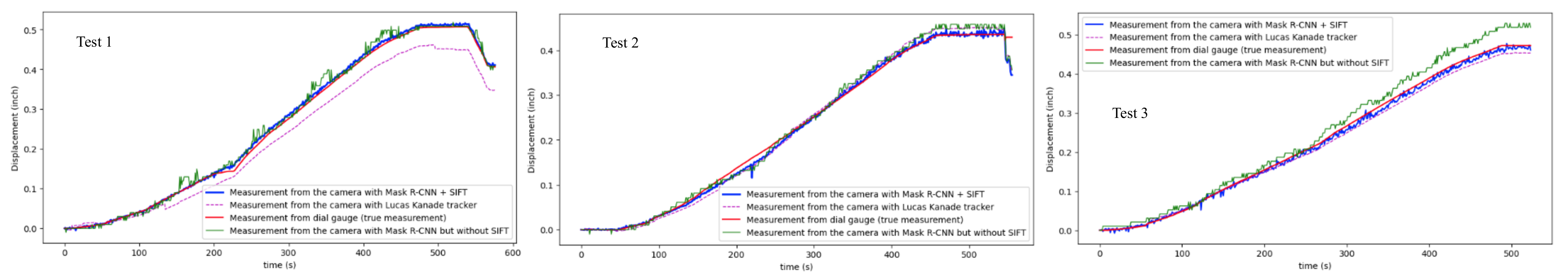}
        \vspace{-8mm}
        \caption{The deflection-time relationship measured by a camera and a dial gauge for the three testing beams.} 
        \label{fig:plot}
        \vspace{-2mm}
\end{figure*}

\begin{figure*}[h!] % h:here, t:top
        \centering
        \includegraphics[width=1\linewidth, height=3.cm]{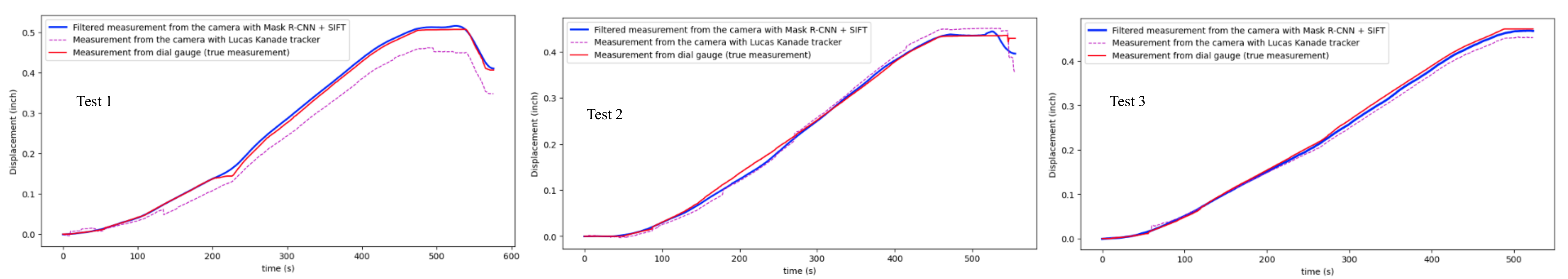}
        \vspace{-8mm}
        \caption{The filtered deflection-time relationship measured by a camera and a dial gauge for the three testing beams.} 
        \label{fig:MAE}
        \vspace{-4mm}
\end{figure*}

The data process for the videos of three tests is like this: since this is a static test, which means the loading and deflection of the tested beams is slow, images from the video at each second are selected as the visual data. There are a total of 500 to 600 images for each test. To train the Mask R-CNN, only 50 images are randomly selected and labeled for detecting and tracking the top of wood frame (in purple mask) as shown in Fig. \ref{fig:maskrcnn}. The LK tracker is used to track the same object and measure the deflection. The testing results are shown in Fig. \ref{fig:plot}. Compared to the LK tracker, this Mask R-CNN with SIFT can provide a measurement closer to the ground truth of beam deflections by a dial gauge. In addition, Savitzky-Golay filter is applied to smooth the measurement such that it becomes consistent (see Fig. \ref{fig:MAE}). Table \ref{tab1} shows MAE (mean absolute error) for both methods. It can be inferred that SIFT and Savitzky-Golay filter can effectively readjust the position of the bounding box predicted from the Mask R-CNN and smooth the measurements, hence, the proposed method can outperform the LK tracker.
\vspace{-4mm}
\begin{table}[htbp]
\caption{MAE for two methods (inches)}
\vspace{-5mm}
\begin{center}
\begin{tabular}{|c|c|c|c|}
\hline
% \textbf{Method}&\multicolumn{3}{|c|}{\textbf{Test er}} \\
\cline{2-4} 
\textbf{Methods} & \textbf{\textit{Test 1}}& \textbf{\textit{Test 2}}& \textbf{\textit{Test 3}} \\
\hline
Mask R-CNN + SIFT & 0.005&0.005 &0.005  \\
LK tracker & 0.030&0.012 &0.012  \\
\hline
\end{tabular}
\label{tab1}
\end{center}
\vspace{-7mm}
\end{table}
\vspace{-1mm}
\subsection{Vibration Measurement of A Shaking Table Test}
\begin{figure}[h!] % h:here, t:top
        \centering
        \includegraphics[width=1\linewidth, height=4.0cm]{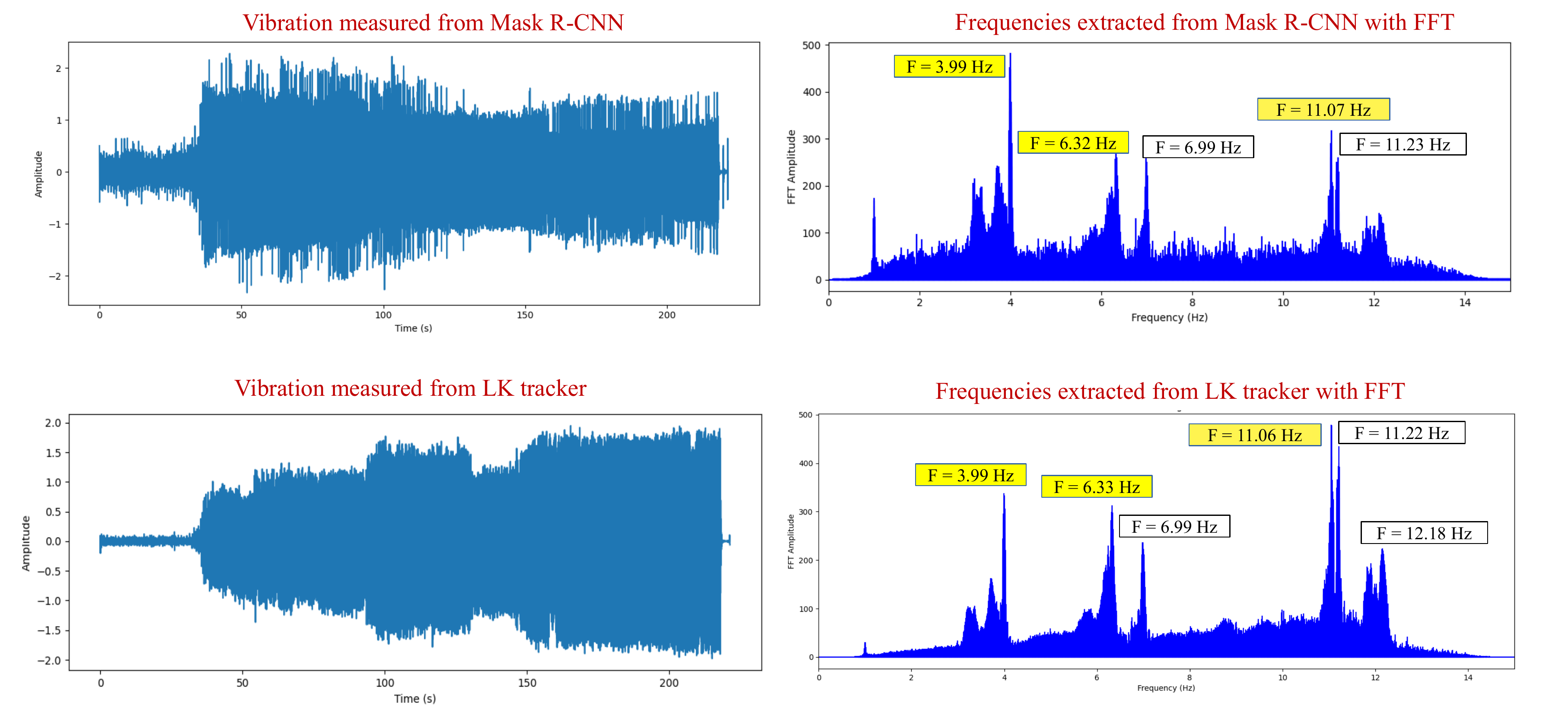}
        \vspace{-8mm}
        \caption{Vibration and frequencies of the shaking table measured by Mask R-CNN and the LK tracker.} 
        \label{fig:table_1}
\end{figure}

\begin{figure}[h!] % h:here, t:top
        \vspace{-8mm}
        \centering
        \includegraphics[width=1\linewidth, height=6.5cm]{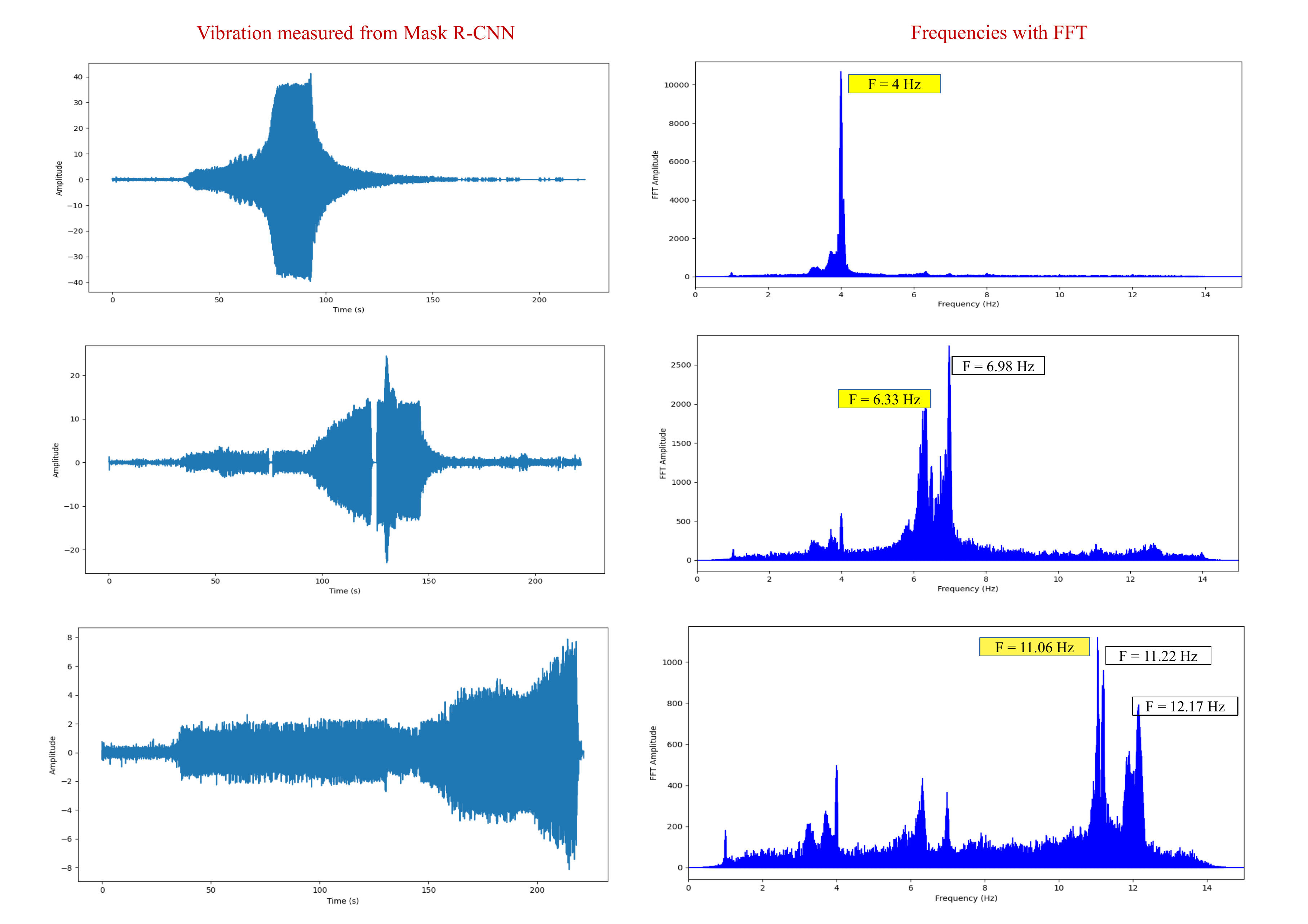}
        \vspace{-8mm}
        \caption{Vibrations of three rectangles (the highest to the lowest one from top to bottom on the left hand side) measured by Mask R-CNN and and the corresponding frequencies calculated by FFT (on the right hand side).} 
        \label{fig:table_2}
        \vspace{-2mm}
\end{figure}

\vspace{-1mm}
Our proposed method was also applied on a shaking table test \cite{quanser_2017} to check its applicability of monitoring dynamic movement of objects. In this test, there are three rectangles (masses) fixed on the shaking table at different heights (see Fig. \ref{fig:shaking}). Each rectangle, which is supported by two sticks, like a structure has its unique resonant frequency in the horizontal direction. This is due to differences between the lateral stiffness of each pair of sticks. The frequencies of the applied shaking are increased from 4 Hz to 13.65 Hz to excite these masses and cause their harmonic vibrations. From the recorded video \cite{quanser_2017}, 150 frames are randomly selected from a total of 6,674 frames and labeled for training the Mask R-CNN. The video has an image size of 640$\times$480 and a frame rate of 30 per second. SIFT is not applied to smooth the measurements here, since the goal of this test is to detect the frequencies instead of accurate amplitudes of the vibration, which is in pixel unit in this test. Thus, the motion of the bounding box represents the translation of each object. The LK tracker is utilized to verify our method by tracking the same vibration of the shaking table. On one hand, all the raw data are processed by Butterworth filter, and FFT is applied to extract the frequencies for each tracking target. The filtered vibrations of the shaking table with the LK tracker and Mask R-CNN are shown in the left figures of Fig. \ref{fig:table_1}. There are three frequencies of vibration at approximately 4 Hz, 6.35 Hz and 11.35 Hz excited by the table. Both methods capture these frequencies (yellow captions in the right figures of Fig. \ref{fig:table_1}) with a less than $2.6\%$ error. On the other hand, the vibrations of three rectangles are measured by the Mask R-CNN and raw data are processed like the procedures for the shaking table. As shown in Fig. \ref{fig:table_2}, their resonant frequencies are very close to the intended frequencies (i.e., 4 Hz, 6.35 Hz and 11.35 Hz). The error rate for this measurement is $0\%$, $0.3\%$ and $2.6\%$, respectively. This indicates that the proposed Mask R-CNN can be used alone to track multiple objects and capture their vibrations characteristics precisely. 
\vspace{-4mm}
% \addtocounter{figure}{1}
% \addtocounter{figure}{8}
% \addtocounter{figure}{-1}
% \addtocounter{figure}{-2}
\vspace{-6mm}
\section{Conclusions}
A deep learning method (i.e., Mask R-CNN with HRNet) and techniques such as SIFT and Savitzky-Golay filter are applied to automatically track the targets and provide the accurate measurement of their motions with a stationary camera. In our first experiment, Mask R-CNN and SIFT were used for precise deflection measurement of the tested RC beams, since SIFT can utilize the keypoints on the targets to refine the measurement. Our method can be closer to the measurement from traditional structural sensors and outperform the LK tracker. The Mask R-CNN was also used alone to track the vibration of multiple targets in a shaking table experiment and capture the resonant frequencies of these targets via Butterworth filter and FFT. These preliminary tests show that the proposed method is robust and has the potential for measuring displacements and vibrations of structural specimens precisely and automatically in laboratory experiments. 

Our ongoing work involves application of the proposed method on actual buildings tested in the field to confirm its applicability for outdoor environments. Other deep learning methods are also being explored and tested.

\vspace{18mm}
\bibliographystyle{ieeetr}
\bibliography{main} 
% \vspace{12pt}
% \color{red}
% IEEE conference templates contain guidance text for composing and formatting conference papers. Please ensure that all template text is removed from your conference paper prior to submission to the conference. Failure to remove the template text from your paper may result in your paper not being published.

\end{document}